\documentclass[aps,pra,preprint,a4paper,showpacs,showkeys,superscriptaddress,nofootinbib]{revtex4-1}

\usepackage{latexsym}
\usepackage{amsmath,amssymb}
\usepackage{graphicx}
\usepackage{subfigure}
\usepackage{xcolor}
\usepackage{physics}
\usepackage{cancel}
\usepackage{tikz}
\usepackage{multirow,tabularx}

\newcolumntype{C}[1]{>{\centering\arraybackslash}p{#1}}
\usepackage{hyperref}     
\hypersetup{colorlinks,%
  citecolor=blue,%
  linkcolor=cyan,%
}

\usepackage[titletoc]{appendix}
\usepackage{enumerate}

\begin{document}


\title{Batalin-Tyutin quantization of dynamical boundary of AdS\texorpdfstring{$_2$}{2}}

\author{Wontae Kim}%
\email[]{wtkim@sogang.ac.kr}%
\affiliation{Department of Physics, Sogang University, Seoul, 04107,
	Republic of Korea}%
\affiliation{Center for Quantum Spacetime, Sogang University, Seoul 04107, Republic of Korea}%

\author{Mungon Nam}%
\email[]{clrchr0909@sogang.ac.kr}%
\affiliation{Department of Physics, Sogang University, Seoul, 04107,
	Republic of Korea}%
\affiliation{Center for Quantum Spacetime, Sogang University, Seoul 04107, Republic of Korea}%
\date{\today}

\begin{abstract}
In a two-dimensional AdS space,
a dynamical boundary of AdS space was described by a one-dimensional quantum-mechanical Hamiltonian
with a coupling between the bulk and boundary system.
In this paper, we present a Lagrangian corresponding to the Hamiltonian through the Legendre transformation
with a constraint.
In Dirac's constraint analysis,
we find two primary constraints without secondary constraints;
however, they are fully second-class.
In order to make the second-class constraint system
into a first-class constraint system, we employ the Batalin-Tyutin Hamiltonian method,
where the extended system reduces to the original one for the unitary gauge condition.
In the spirit of the AdS/CFT correspondence,
it raises a question whether
a well-defined extended bulk theory corresponding to the extended boundary theory
can exist or not.

\end{abstract}

%



\maketitle


\raggedbottom

\section{Introduction}
\label{sec:introduction}
Some decades ago, Jackiw and Teitelboim (JT) proposed a two-dimensional model for anti-de Sitter (AdS) space  \cite{Jackiw:1984je,Teitelboim:1983ux}. Recently, Almheiri and Polchinski (AP)
also studied a modified model allowing one to set up more meaningful holographic dictionary
through analysis of the boundary dynamics~\cite{Almheiri:2014cka}.
Since the holographic AdS/CFT dictionary is still incomplete, such lower dimensional examples amenable to our needs
would provide an essential feature of AdS/CFT correspondence.
In this regard, Engels\"oy,  Mertens, and Verlinde studied the black hole evaporation process from AdS$_2 $ holography~\cite{Engelsoy:2016xyb}.
One of the interesting conclusions is that
the time coordinate becomes dynamical on the boundary, and
the one-dimensional boundary action is given by the Schwarzian derivative \cite{Maldacena:2016upp}. They also obtained
the relevant quantum-mechanical Hamiltonian by taking into account a coupling between the bulk and the boundary.

The above Hamiltonian for the dynamical boundary of AdS$_2$
can be translated into a corresponding Lagrangian through the
Legendre transformation. Then, we immediately find two primary constraints from the definition
of momenta, but they are unexpectedly second-class without secondary constraints
when classified by the Dirac method \cite{dirac2001lectures}.
What second-class constraints imply is that a local symmetry
implemented by constraints as symmetry generators would be broken.
If the second-class constraint system is converted into a first-class one,
then the remaining quantization process will follow ordinary methods in Refs.~\cite{Fradkin:1975cq,Henneaux:1985kr,Becchi:1975nq,Kugo:1979gm,Batalin:1986fm,Batalin:1991jm}.

In fact, there are largely two ways to realize first-class constraint systems.
The first one is to use the action. Faddeev and Shatashivili~\cite{Faddeev:1986pc}
introduced the Wess-Zumino action~\cite{Wess:1971yu} in order to cancel out the gauge anomaly
responsible for second-class constraint algebra
and they eventually obtained the first-class constraint algebra.
Subsequently, the Wess-Zumino action to cancel the gauge anomaly was also studied in Refs.~\cite{Babelon:1986sv,Harada:1986wb,Miyake:1987dk}.
The second one is to use the Hamiltonian formalism such as Batalin-Tyutin Hamiltonian method~\cite{Batalin:1991jm}.
Interestingly, Banerjee~\cite{Banerjee:1993pm} applied the method to the Chern-Simons field theory of second-class constraint system and obtained a strongly involutive constraint algebra in an extended phase space, which yields
a new Wess-Zumino type action which cannot be derived from the action level. And its non-Abelian extension was also done
in Ref.~\cite{Kim:1994np}.
In addition, the method was applied to wide variety of cases of interest: anomalous gauge theory~\cite{Fujiwara:1989ia,Kim:1992ey,Banerjee:1993pj,Fujiwara:1990rx},
non-gauge theories~\cite{Hong:1999gx,Hong:2000bp,Hong:2000ex}, and chiral bosons \cite{Amorim:1994ft,Amorim:1994np,Kim:2006za}.

In this paper, we will consider a Lagrangian describing the dynamical boundary of AdS$_2$ compatible with the Hamiltonian in Ref.~\cite{Engelsoy:2016xyb}. Then,
we obtain two primary constraints from the definition of momenta;
however, they turn out to be second-class constraints.
Thus, we would like to study how to get the first-class constraint system
by the use of the Batalin-Tyutin Hamiltonian method.
The organization of the paper is as follows.
In Sec.~\ref{sec:Preliminaries}, we will recapitulate
the derivation of the Hamiltonian for the dynamical boundary of AdS$_2$ from the AP model.
In Sec.~\ref{sec:Hamiltonian formalism}, we obtain the corresponding Lagrangian to the Hamiltonian studied
in Sec.~\ref{sec:Preliminaries}.
The two primary constraints are shown to be second-class so that
the Lagrange multipliers are fully determined without any further secondary constraints.
In Sec.~\ref{sec:Batalin-Tyutin quantization}, using the Batalin-Tyutin Hamiltonian method,
we realize the first-class constraint system and obtain the involutive Hamiltonian.
Finally, a conclusion will be given in Sec.~\ref{sec:conclusion}.

\section{Holographic renormalization of the AP model}
\label{sec:Preliminaries}
In the AP model \cite{Almheiri:2014cka}, we encapsulate the derivation of the Hamiltonian describing
the dynamical boundary of AdS$_2$ from a holographic renormalization process \cite{Almheiri:2014cka,Engelsoy:2016xyb}.
The action is given as
\begin{align}
	S&= S_{\rm AP} + S_{\rm GH} + S_{\rm matt},\label{eq:total action}\\
	S_{\rm AP} &= \frac{1}{16\pi G}\int \dd^2x\sqrt{-g} [\Phi^2(R+2)-2],\label{eq:JT action}\\
		S_{\rm GH} &= \frac{1}{8\pi G}\int \dd t\sqrt{-\gamma}\Phi^2 K,\label{eq:boundary term}
\end{align}
where $S_{\rm AP}$ is the AP action \cite{Almheiri:2014cka}, $S_{\rm GH}$ is the Gibbons-Hawking term
\cite{Gibbons:1976ue}, $S_{\rm matt}$ is some arbitrary matter system coupled to the two dimensional metric
$\dd s^2=g_{\mu \nu}\dd x^\mu \dd x^\nu$, and $ \Phi $ is a dilaton.
In the conformal gauge, the length element is written as $ \dd s^2 = -e^{2\omega(X^+,X^-)}\dd X^+\dd X^-$
where $X^{\pm} = X^0 \pm X^1 = T\pm Z$. From the action \eqref{eq:total action}, equations of motion are derived as
\begin{align}
	 & \partial_+\partial_-\omega + \frac{1}{4}e^{2\omega} = 0,\label{eq:eom metric}\\
	 & \partial_+\partial_-\Phi^2 + \frac{1}{2}e^{2\omega}(\Phi^2-1) = 8\pi GT_{+-},\label{eq:eom dilaton}\\
	 & -e^{2\omega}\partial_{\pm}(e^{-2\omega}\partial_{\pm}\Phi^2) =
                                  8\pi G T_{\pm\pm},\label{eq:constraints}
\end{align}
where the stress tensor for matter is $T_{\mu\nu} = -(2/\sqrt{-g})\delta{S_{\rm matt}}/ \delta{g^{\mu\nu}}$.
The general solution to Eq.~\eqref{eq:eom metric} is obtained as the AdS$_{2}$ geometry
\begin{equation}
\label{Poincare metric}
	e^{2\omega} = \frac{4}{(X^+ - X^-)^2}.
\end{equation}
For $N$ conformal fields of $ T_{+-} = 0 $,
the dilaton solution takes the following form:
\begin{equation}
\label{eq:dilaton sol}
	\Phi^2 = 1 + \frac{a - 8\pi G(I_+ + I_-)}{X^+ - X^-},
\end{equation}
where integrated source terms are given by
$I_+ = \int_{X^+}^{\infty}\dd s(s-X^+)(s-X^-)T_{++}(s)$ and $I_- = \int_{-\infty}^{X^-}\dd s(s-X^+)(s-X^-)T_{--}(s)$.

If the conformal matter fields received quantum corrections, then the stress tensor for matter would be no longer traceless.
In our work, we will focus on the classically conformal matter fields
without trace anomaly.
The integration constant $a$ in Eq.~\eqref{eq:dilaton sol}
is assumed to be positive, which prevents strong coupling singularity
from reaching the boundary \cite{Almheiri:2014cka}.
In particular, the infalling source is assumed to be $ T_{++} = 0 $ and $ T_{--} = E\delta(s) $,
then Eq.~\eqref{eq:dilaton sol} can be explicitly expressed by
\begin{equation}\label{eq:black hole sol}
	\Phi^2 = 1+a\frac{1 - \kappa E X^+X^-\Theta(X^-)}{X^+ - X^-},
\end{equation}
where $\kappa = 8\pi G/a$ and $\Theta$ is a step function.
Note that the Poincar\'e vacuum and the massive black hole are
characterized by dilaton profile for $X^-<0$ and for $X^->0$, respectively.
For the latter case, the black hole can be expressed in terms of a static form by the use of the coordinate transformations
$X^{\pm}(\sigma^{\pm}) = ({1/\sqrt{\kappa E}})\tanh(\sqrt{\kappa E}\sigma^{\pm})$,
where $\sigma^{\pm}=t \pm \sigma$.
Then, the future and the past horizon are found at
$X^+( \infty ) \to 1/\sqrt{\kappa E}$ and
$X^-(-\infty)\to -1/\sqrt{\kappa E}$, respectively.
In addition, the unperturbed boundary is located at $X^+(\sigma^+) = X^-(\sigma^-)$ which is assumed to be coincident
with $\sigma^+ = \sigma^-$, {\it i.e.}, $\sigma^1 =0$.
Then, a dynamical boundary time is naturally defined as $X^+(t) = X^-(t) = T(t)$.

Using an infinitesimally small cut-off $\epsilon$ from the unperturbed boundary,
one can define two quantities such as
$ X^+(t+\epsilon) + X^-(t-\epsilon)  = 2T(t)$ and
	$ X^+(t+\epsilon) - X^-(t-\epsilon)  =  2\epsilon \dot{T}(t)$.
In fact, the essential requirement in Ref.~\cite{Engelsoy:2016xyb} is that the asymptotic behaviour of the
dilaton is the same as that of the Poincar\'e patch at the boundary so that
\begin{equation}\label{eq:require dilaton}
	\Phi^2(t) = \frac{a}{2\epsilon \dot{T}}\left[ 1-\kappa(I_+(t) + I_-(t)) \right] = \frac{a}{2\epsilon},
\end{equation}
which dictates $\dot{T}(t) = 1-\kappa(I_+(t) + I_-(t))$.
Hence, the equation of motion for the dynamical boundary can be obtained as
\begin{equation}
\label{eq:eom dyn bdy time}
	\frac{1}{2\kappa}\dv[2]{t}\log\dot{T}+(P_+ - P_-)\dot{T} = 0,
\end{equation}
where $P_+ = \int_{T}^{\infty}\dd sT_{++}(s)$ and $P_- = -\int_{-\infty}^{T}\dd s T_{--}(s)$.
Using Eq.~\eqref{eq:eom dyn bdy time}, near the boundary, one can get the metric \eqref{Poincare metric} and the dilaton \eqref{eq:dilaton sol} as
\begin{align}
	e^{2\omega} = \frac{1}{\epsilon^2} + \frac{2}{3}\{ T,t \} + \mathcal{O}(\epsilon^2),\quad
	\Phi^2= \frac{a}{2\epsilon} + 1 - \frac{a}{3}\{ T,t \} + \mathcal{O}(\epsilon^2),\label{eq:bdy dilaton}
\end{align}
where
$\{ T,t \} = \dddot{T}/\dot{T} - (3/2) ( \ddot{T}/\dot{T} )^2$ is the Schwarzian derivative.

Let us now get a boundary stress tensor of the dual CFT through
the holographic renormalization procedure~\cite{Almheiri:2014cka,Engelsoy:2016xyb}. Varying the renormalized
on-shell action of
$S_{\rm ren} = S_{\rm AP} + S_{\rm ct}$
including a counter term
$S_{\rm ct} = 1/(8\pi G)\int\dd t\sqrt{-\gamma}(1-\Phi^2)$
with respect to the boundary metric $\hat{\gamma}_{tt}$, one can obtain
\begin{equation}
\label{eq:bdy stress tensor def}
	\langle \hat{T}_{tt} \rangle = -\frac{2}{\sqrt{-\hat{\gamma}}}\fdv{S_{\rm ren}}{\hat{\gamma}^{tt}} = \lim_{\epsilon\to 0}-\frac{2\epsilon}{\sqrt{-\gamma(\epsilon)}}\fdv{S_{\rm ren}(\epsilon)}{\gamma^{tt}(\epsilon)},
\end{equation}
where $ \hat{\gamma}^{tt} = \lim_{\epsilon\to 0}\gamma^{tt}/\epsilon^2 $ is the metric of the boundary. Thus, the boundary stress tensor
is obtained as
\begin{eqnarray}
\label{eq:bdy stress tensor}
    \langle \hat{T}_{tt} \rangle &=& \frac{\epsilon}{8\pi G}[e^{\omega}\partial_{\epsilon}\Phi^2 - e^{2\omega}(1-\Phi^2) ]\\
	 &=& -\frac{1}{2\kappa}\{ T(t),t \},
\end{eqnarray}
by plugging Eq.~\eqref{eq:bdy dilaton} into Eq.~\eqref{eq:bdy stress tensor}.
The boundary stress tensor $\langle \hat{T}_{tt} \rangle$
reflects excitations of the boundary in terms of the static time coordinate $t$
and it must be a Hamiltonian for the dynamical boundary.
In order to describe the coupling between the matter sector and the
dynamical boundary theory,
the authors in Ref.~\cite{Engelsoy:2016xyb} introduced a new variable $\varphi =\log \dot{T}$ and then arrived at
the Hamiltonian
\begin{equation}
\label{EMV}
H_{\rm EMV} = \kappa\pi_{\varphi}^2 + \pi_{T} e^{\varphi}+e^{\varphi}(P_+-P_-),
\end{equation}
where $\pi_{\varphi}$ and $\pi_{T}$ are conjugate momenta corresponding to $\varphi$ and $T$, respectively,
and the Hamiltonian reduces to $H_{\rm EMV} = \kappa\pi_{\varphi}^2+e^{\varphi}(P_+-P_-)$
upon setting $\pi_{T}=0$.
\section{Hamiltonian formulation of the dynamical boundary}
\label{sec:Hamiltonian formalism}
We derive a Lagrangian corresponding to the Hamiltonian~\eqref{EMV} by means of the canonical path-integral.
Thus, we consider the partition function as
\begin{align}
	\mathcal{Z} &= \int\mathcal{D}\varphi\mathcal{D}T\mathcal{D}\pi_{\varphi}\mathcal{D}\pi_{T}\exp[i\int\dd t (\pi_{\varphi}\dot{\varphi} + \pi_{T}\dot{T} - H_{\rm HMV})]\nonumber\\
				&= \int\mathcal{D}\varphi\mathcal{D}T\delta(\dot{T}- e^{\varphi})\exp[i\int\dd t \left( \frac{1}{4\kappa}\dot{\varphi}^2 - e^{\varphi}(P_+ - P_-)  \right)]\nonumber\\
				&= \int\mathcal{D}\varphi\mathcal{D}T\mathcal{D}\lambda\exp[i\int\dd t \left( \frac{1}{4\kappa}\dot{\varphi}^2 + \lambda(\dot{T}- e^{\varphi}) - e^{\varphi}(P_+ - P_-)  \right)].
\label{eq:partition funcs}
\end{align}
In Eq.~\eqref{eq:partition funcs}, the integration with respect to $\pi_{T}$ gives the delta functional,
which is rewritten in terms of a new variable $\lambda$ in the last line.
Thus, we obtain the action describing the dynamical boundary of AdS$_{2}$ as
\begin{equation}
\label{eq:bdy dyn action}
	S = \int\dd t L= \int\dd t \left( \frac{1}{4\kappa}\dot{\varphi}^2 + \lambda(\dot{T}-e^{\varphi}) - e^{\varphi}(P_+ - P_-)  \right),
\end{equation}
where $\lambda$ plays the role of Lagrange multiplier to implement $\dot{T}-e^{\varphi}=0$ and
$P_\pm$ are assumed to be background sources.

The canonical momenta conjugate to variables $(\varphi, T, \lambda)$ are defined as
\begin{equation}
\label{eq:canonical momenta}
	\pi_{\varphi} = \frac{1}{2\kappa}\dot{\varphi},\quad \pi_{T} = \lambda,\quad \pi_{\lambda}=0.
\end{equation}
Then, the canonical Hamiltonian is obtained
through the Legendre transformation of the Lagrangian \eqref{eq:bdy dyn action} as
\begin{align}
\label{eq:canonical H}
	H_{\rm c} = \kappa\pi_{\varphi}^2 + \lambda e^{\varphi}+e^{\varphi}(P_+-P_-),
\end{align}
which is the same as Eq.~\eqref{EMV} as it must be when $\lambda$ is replaced by $\pi_{T}$.
The standard Poisson brackets are imposed as follows,
\begin{equation}
\label{eq:canonical rel}
	\{ \varphi,\pi_{\varphi} \}_{\rm PB} = 1,\quad \{ T,\pi_{T} \}_{\rm PB} = 1,
\quad \{ \lambda,\pi_{\lambda} \}_{\rm PB} = 1.
\end{equation}
In Eq.~\eqref{eq:canonical momenta},
we now identify two primary constraints as \cite{dirac2001lectures}
\begin{equation}\label{eq:bdy constraints}
	\Omega_1 = \pi_{T} - \lambda \approx 0 ,\quad \Omega_2 = \pi_{\lambda} \approx 0,
\end{equation}
and then construct the primary Hamiltonian by adding the two primary constraints to the canonical Hamiltonian as
\begin{equation}
\label{eq:primary H}
	H_{\rm p} = H_{\rm c} + u_1\Omega_1 + u_2\Omega_2,
\end{equation}
where $u_1$ and $u_2$ are arbitrary Lagrange multipliers.
The stability of primary constraints with respect to time evolution is
\begin{align}
	\{ \Omega_1,H_{\rm p} \}_{\rm PB} &= -u_2 \approx 0,\label{eq:first pb}\\
	\{ \Omega_2,H_{\rm p} \}_{\rm PB} &= -e^{\varphi} + u_1 \approx 0.\label{eq:second pb}
 \end{align}
Note that the two Lagrange multipliers can be chosen as $u_1 =e^{\varphi}$ and $u_2 =0$,
and thus, they are fully fixed since the primary constraints are second-class.
Accordingly, the Dirac bracket between canonical variables can be defined as
\begin{equation}\label{eq:dirac bracket def}
	\{ A, B \}_{\rm D} = \{ A,B \}_{\rm PB} - \sum_{i,j}^{2}\{ A,\Omega_i \}_{\rm PB}C^{-1}_{ij}\{ \Omega_j,B \}_{\rm PB},
\end{equation}
where the Dirac matrix is $C_{ij} = \{ \Omega_i,\Omega_j \}_{\rm PB} =-\epsilon_{ij}$ with
$\epsilon_{12}=1$.
Hence, the non-vanishing Dirac brackets are
\begin{equation}\label{eq:dirac bracket}
	\{ \varphi,\pi_{\varphi} \}_{\rm D} = 1,\quad \{ T,\pi_{T} \}_{\rm D} = 1,\quad \{ \lambda,T \}_{\rm D} = -1.
\end{equation}
The equations of motion are given as
\begin{align}
	\dot{\varphi}=\{ \varphi,H_{\rm p} \}_{\rm D} = 2\kappa\pi_{\varphi},&\quad \dot{\pi}_{\varphi} = \{ \pi_{\varphi},H_{\rm p} \}_{\rm D} =  -\lambda e^{\varphi}-e^{\varphi}(P_+-P_-),\label{eq:phi eom}\\
	\dot{T} = \{ T,H_{\rm p} \}_{\rm D} = e^{\varphi},&\quad \dot{\pi}_{T} = \{ \pi_{T},H_{\rm p} \}_{\rm D} =  0.\label{eq:T eom}
\end{align}
In Eq.~\eqref{eq:phi eom}, the combined first-order differential equations result in
\begin{equation}\label{eq:phi equ}
	\frac{1}{2\kappa}\ddot{\varphi} + (\lambda +P_+ -P_-) e^{\varphi} = 0.
\end{equation}
In Eq.~\eqref{eq:T eom}, $ \pi_{T} $ turns out to be constant so that $ \lambda $ must also be constant
through the constraint $\Omega_1$.
For simplicity, if we set $\lambda =0$, then Eq.~\eqref{eq:phi equ} is identical with
the dynamical equation of motion \eqref{eq:eom dyn bdy time}
and the reduced Hamiltonian is
$H_r = \kappa\pi^2_{\varphi} + e^{\varphi} (P_+ - P_-)$.

Consequently, the boundary system dual to the bulk AdS$_2$ is found to be the second-class constraint system.
Thus, we will make the second-class system to the first-class one in virtue of the Batalin-Tyutin Hamiltonian
embedding.
\section{Batalin-Tyutin Hamiltonian quantization}
\label{sec:Batalin-Tyutin quantization}
Following the Batalin-Tyutin quantization
method \cite{Batalin:1991jm}, we introduce new auxiliary one-dimensional fields
$\theta^1$ and $\theta^2$ in order to convert second-class constraints into first-class ones
in the extended phase space. Let us assume the Poisson algebra
satisfying
\begin{equation}\label{eq:aux fields}
	\{ \theta^i,\theta^j \}_{\rm PB} = \omega^{ij}.
	\end{equation}
In the extended phase space, the modified constraints $\tilde{\Omega}_i$ are assumed to be \cite{Batalin:1991jm}
\begin{equation}\label{eq:mod const def}
	\tilde{\Omega}_i = \Omega_i + \sum_{n=1}^{\infty}\Omega_i^{(n)},\quad \Omega^{(n)}_i \sim (\theta^i)^n,
\end{equation}
with the boundary condition $ \tilde{\Omega}_i = \Omega_i $ for $ \theta^i = 0 $.
The first order correction is
\begin{equation}\label{eq:1st corr cost}
	\Omega^{(1)}_i = X_{ij}\theta^j
\end{equation}
for some matrix $ X_{ij}$, and thus, we require
\begin{equation}\label{eq:mat equ}
	\{ \tilde{\Omega}_i,\tilde{\Omega}_j \}_{\rm PB} = C_{ij} + \{ X_{ik}\theta^k,X_{lj}\theta^l \}_{\rm PB} = C_{ij} + X_{ik}\omega^{kl}X_{lj} = 0.
\end{equation}
As was emphasized in Ref.~\cite{Banerjee:1993pm}, there exists an arbitrariness
in choosing $\omega_{ij}$ and $X_{ij}$, which corresponds to canonical transformation in the
extended phase space \cite{Batalin:1986fm,Batalin:1991jm}.
Now, the simplest choice for $\omega^{ij}$ and $ X_{ij} $ is
\begin{equation}\label{eq:mat X}
\omega^{ij}=-\epsilon^{ij}, \quad	X_{ij} = \delta_{ij},
\end{equation}
where $\epsilon^{12}=-1$.
Thus, the modified constraints in the extended phase space are given as
\begin{equation}\label{eq:mod consts}
	\tilde{\Omega}_1 = \pi_{T} - \lambda + \theta^1,\quad \tilde{\Omega}_2 = \pi_{\lambda} + \theta^2
\end{equation}
without any higher order corrections.

On the other hand, in the extended phase space,
the involutive Hamiltonian is defined by \cite{Batalin:1991jm}
\begin{equation}
\label{eq:invol H def}
	\tilde{H} = H_c + \sum_{n=1}^{\infty}H^{(n)},
\end{equation}
with the $ n $th order correction $ H^{(n)} $ being
\begin{equation}
\label{eq:nth corr H}
	H^{(n)}=-\frac{1}{n}\theta^i\omega_{ij}X^{jk}G_{k}^{(n-1)} \quad (n\geq 1),
\end{equation}
where $\omega_{ij}$ and $ X^{ij}$ are inverse matrices of $\omega^{ij}$ and $ X_{ij}$, respectively.
The generating functions $ G^{(n)}_a $ are given as
\begin{align}
	G_{i}^{(0)} &= \{ \Omega_i, H_c \}_{\rm PB},\label{eq:0th gen func def}\\
	G_{i}^{(n)} &= \{ \Omega_i,H^{(n)} \}_{\rm PB,\mathcal{O}} + \{ \Omega^{(1)}_a,H^{(n-1)} \}_{\rm PB, \mathcal{O}},
\quad (n\geq 1),\label{eq:nth gen func def}
\end{align}
where $ \mathcal{O} $ implies that the given Poisson brackets are calculated with respect to the original variables.
Explicitly, they are obtained as
\begin{equation}\label{eq:gen funcs}
	G^{(0)}_1 = -e^{\varphi},\quad G^{(0)}_2 = 0,\quad G^{(n)}_i = 0, \quad ( n\geq 1).
\end{equation}
Hence, there only exists a linear order correction so that
\begin{equation}\label{eq:1st corr H}
	H^{(1)} = -e^{\varphi}\theta^1,\quad  H^{(n)} = 0,\quad (n\geq 2).
\end{equation}
The final expression for the involutive Hamiltonian~\eqref{eq:invol H def} is given as
\begin{equation}\label{eq:invol H}
	\tilde{H} = H_c + H^{(1)} = \kappa\pi_{\varphi}^2 +e^{\varphi}(\lambda - \theta^1) +e^{\varphi}(P_+-P_-),
\end{equation}
and the time evolution of the modified constraints does not generate any more constraints since
\begin{equation}\label{eq:stab involv H}
	\{ \tilde{\Omega}_i,\tilde{H} \}_{\rm PB} = 0.
\end{equation}
In the above analysis for the constraint system, we see that the original second-class constraint
system can be converted into the first-class system by introducing two auxiliary fields
in the extended phase space.

Next, let us consider the phase space partition function,
\begin{equation}
\label{eq:partition func def}
	\mathcal{Z} = \int\mathcal{D}\varphi\mathcal{D}T\mathcal{D}\lambda\mathcal{D}\pi_{\varphi}\mathcal{D}\pi_{T}\mathcal{D}
\pi_{\lambda}\mathcal{D}\theta^1\mathcal{D}\theta^2\prod_{i,j}^{2}\delta(\tilde{\Omega}_i)\delta(\Gamma_j)
\left|\det(\tilde{\Omega}_i,\Gamma_j)\right|e^{iS'},
\end{equation}
where $S' = \int \dd t (\pi_{\varphi}\dot{\varphi} + \pi_{T}\dot{T} + \pi_{\lambda}\dot{\lambda} + \theta^{2}\dot{\theta}^1 - \tilde{H})$ and $\Gamma_i$ are gauge conditions.
Integrating out the momenta of $ \pi_{\varphi}, \pi_{T},\pi_{\lambda} $ in Eq.~\eqref{eq:partition func def}, we finally get
\begin{eqnarray}
\label{eq:extend L}
\tilde{S}&=&S +S_{\rm WZ},\\
S_{\rm WZ}&=& \int\dd t\left( -\theta^1(\dot{T}-e^{\varphi}) - \theta^2(\dot{\lambda} - \dot{\theta}^1)\right) \label{WZ}.
\end{eqnarray}
Expectedly, if we choose the unitary gauge condition as
$\Gamma_i=\theta^i =0$, the extended theory reduces to the original action \eqref{eq:bdy dyn action}.
The action \eqref{WZ} is the new type of Wess-Zumino action derived from
the Batalin-Tyutin Hamiltonian formalism.
Note that
this formalism, embedding of familiar second-class constraint systems
~\cite{Fujiwara:1989ia,Kim:1992ey,Banerjee:1993pj,Fujiwara:1990rx} reproduces the Wess-Zumino action derived from usual Lagrangian procedures
\cite{Babelon:1986sv,Harada:1986wb,Miyake:1987dk}.
The present Batalin-Tyutin formulation is one of the different applications to non-gauge theories studied in Refs.~\cite{Hong:1999gx,Hong:2000bp,Hong:2000ex,Amorim:1994ft,Amorim:1994np,Kim:2006za}, so
the origin of the second-class nature is not the result of a genuine gauge symmetry breaking.

Now, let us elaborate gauge conditions and discuss the role of the auxiliary fields $\theta^1$ and $\theta^2$.
The total action \eqref{eq:extend L} which consists of the original action and the Wess-Zumino action
can be neatly rewritten as
\begin{equation}\label{total}
	\tilde{S} = \int\dd t \left( \frac{1}{4\kappa} \dot{\varphi}^2 - e^{\varphi}(P_+-P_-) + (\lambda-\theta^1)(\dot{T} + \dot{\theta^2} - e^{\varphi}) \right).
\end{equation}
Note that the modified constraints \eqref{eq:mod consts} as symmetry generators indicate that the action \eqref{total}
is invariant under the following local transformations
\begin{align}
	\delta_{\epsilon_1}\theta^1 = \epsilon_1(t),&\quad \delta_{\epsilon_1}\lambda = \epsilon_1(t),\\
	\delta_{\epsilon_2}\theta^2 = \epsilon_2(t),&\quad \delta_{\epsilon_{2}}T = -\epsilon_{2}(t)
\end{align}
implemented by a local symmetry generator
\begin{equation}
\label{}
  Q = (\pi_T - \lambda + \theta^1)\epsilon_2(t) - (\pi_\lambda + \theta^2)\epsilon_1(t),
\end{equation}
where $\epsilon_1(t)$ and $\epsilon_2(t)$ are arbitrary local parameters.
The auxiliary fields
can be eliminated by choosing gauge conditions.
Choosing special local parameters,
one can take an unitary gauge condition such as $\Gamma_1 =\theta^1=0$ and $\Gamma_2 =\theta^2=0$, which results in the original action \eqref{eq:bdy dyn action}.
In that sense, the auxiliary degrees of freedom are gauge artefacts in this special gauge.
However, thanks to the local symmetry, a different kind of gauge condition, for example,
$\Gamma_1 =\lambda=0$ and $\Gamma_2 =T=0$ can be chosen. Then,
the reduced action becomes
\begin{equation}
S = \int\dd t \left( \frac{1}{4\kappa} \dot{\varphi}^2 - e^{\varphi}(P_+-P_-)
-\theta^1(\dot{\theta^2} - e^{\varphi}) \right),
\end{equation}
where physical contents are the same as those of the original action~\eqref{eq:bdy dyn action} because $\theta_1$ and $\theta_2$ play the role
of $\lambda$ and $T$. Consequently, the original theory turns out to be the gauge fixed version of the extended theory.

On the other hand,
the AdS/CFT correspondence tells us that the classical action on the gravity side is the quantum effective action for
the dual conformal theory on the boundary.
In the present JT model coupled to matter, the boundary theory could be described by
the quantum-mechanical Hamiltonian \eqref{EMV}. However, the Hamiltonian system on the boundary was found to belong to
the second-class constraint system which would indicate that a certain local symmetry for the boundary theory was broken.
In order to retrieve the broken local symmetry, the auxiliary
degrees of freedom were added without changing net physical degrees of freedom in the Hamiltonian.
In the spirit of the AdS/CFT correspondence, one might ask what the
extended bulk theory corresponding to the extended boundary theory
\eqref{eq:extend L} is.
The total bulk system may have a richer symmetry-structure compared to that of the original AdS$_{2}$.
This issue was unsolved.

\section{conclusion}
\label{sec:conclusion}
In conclusion,
the dynamical boundary of the two-dimensional AdS space, described by the one-dimensional Hamiltonian
having a coupling between the bulk and boundary system,
we obtained the Lagrangian corresponding to the Hamiltonian.
In Dirac's constraint analysis, there were two primary constraints which are fully second-class.
In order to convert the second-class constraint system
into the first-class constraint system, we employed the Batalin-Tyutin Hamiltonian method
and obtained the closed constraint algebra in the extended space.
The extended system, of course, reduces to the original one for the unitary gauge condition.
From the viewpoint of the AdS/CFT correspondence,
it raises a question regarding the existence of
the extended bulk gravity corresponding to the extended boundary theory,
which deserves further study.

As a comment,
the Hamiltonian \eqref{EMV} reproduces the equation of motion for the dynamical boundary of AdS$_2$
\eqref{eq:eom dyn bdy time} upon setting $\pi_T=0$ \cite{Engelsoy:2016xyb}.
In the Dirac method also, Eq.~\eqref{eq:eom dyn bdy time} was obtained by setting
$\lambda=0$ in Eq.~\eqref{eq:canonical H}, which is actually the same condition as $\pi_T=0$ because they are related to each
other through the constraint \eqref{eq:bdy constraints}.
Thus, one might wonder how about introducing additional constraint to enforce $\lambda=0$ by
means of new auxiliary $\lambda_1$. Now, we can add $\lambda \lambda_1$ to the starting action
\eqref{eq:bdy dyn action}.
After some tedious calculations, we find that $\lambda_1$ still exists in the final equation of motion, so we need to introduce additional term $\lambda_1 \lambda_2$ to enforce $\lambda_1=0$.
Unfortunately, the repeated infinite process would not warrant the condition $\lambda=0$.
We hope this issue will be addressed elsewhere.


\acknowledgments
This work was supported by the National Research Foundation of Korea(NRF) grant funded by the Korea government(MSIT) (No. NRF-2022R1A2C1002894) and
Basic Science Research Program through the National Research Foundation of Korea(NRF) funded by the Ministry of Education through the Center for Quantum Spacetime (CQUeST) of Sogang University (NRF-2020R1A6A1A03047877).


\bibliographystyle{JHEP}       

\bibliography{reference}

\end{document}